\documentclass[conference]{IEEEtran}
\IEEEoverridecommandlockouts

\usepackage{amsfonts}
\usepackage{tabu}
\usepackage{booktabs}
\usepackage{float}
\usepackage{gensymb}
\usepackage{textcomp}
\usepackage{dblfloatfix}
\usepackage{amsmath}
\usepackage{algorithm}
\usepackage{algpseudocode}
\usepackage{pifont}
\usepackage{verbatim}
\usepackage{float}
\usepackage{dblfloatfix}
\usepackage{graphicx}
\usepackage{seqsplit}
\usepackage{caption}
\usepackage{enumitem}
\usepackage{array}
\usepackage{subcaption}
\usepackage{url}

\begin{document}

%\title{Paper Title*\\
\title{Towards a Practical Pedestrian Distraction Detection Framework using Wearables
%\thanks{Research reported in this publication was supported by the United States National Science Foundation (NSF) under award number 1637290.}
}

\author{\IEEEauthorblockN{Nisha Vinayaga-Sureshkanth, Anindya Maiti, Murtuza Jadliwala, \\ Kirsten Crager, Jibo He}
\IEEEauthorblockA{\textit{Wichita State University,}\\ Kansas, USA}
\and
\IEEEauthorblockN{Heena Rathore}
\IEEEauthorblockA{\textit{Qatar University,}\\ Qatar}
}

\maketitle

\begin{abstract}
Pedestrian safety continues to be a significant concern in urban communities and pedestrian distraction is emerging as one of the main causes of grave and fatal accidents involving  pedestrians.
%While distracted driving has been one of the main contributing factors so far, 
%safety related incidents that often result in grave injuries and/or deaths. 
The advent of sophisticated mobile and wearable devices, equipped with high-precision on-board sensors capable of measuring fine-grained user movements and context, provides a tremendous opportunity for designing effective pedestrian safety systems and applications.
% for improving  in these communities.
Accurate and efficient recognition of pedestrian distractions in real-time given the memory, computation and communication limitations of these devices, however, remains the key technical challenge in the design of such systems.
% is: is opportunity, however, also comes with the additional challenge of designing systems that can  
%s in the form of resource constraints and communication limitations.
%In order for any pedestrian distraction detection and safety system to be useful, and to be successfully adopted by end-users, it needs to not only accurately recognize distractions, but also be computationally fast and energy efficient. 
Earlier research efforts in pedestrian distraction detection using data available from mobile and wearable devices have primarily focused only on achieving high detection accuracy, resulting in designs that are either resource intensive and unsuitable for implementation on mainstream mobile devices, or computationally slow and not useful for real-time pedestrian safety applications, or require specialized hardware and less likely to be adopted by most users. In the quest for a pedestrian safety system that achieves a favorable balance between computational efficiency, detection accuracy, and energy consumption, this paper makes the following main contributions: (i) design of a novel complex activity recognition framework which employs motion data available from users' mobile and wearable devices and a lightweight frequency matching approach to accurately and efficiently recognize complex distraction related activities, and (ii) a comprehensive comparative evaluation of the proposed framework with well-known complex activity recognition techniques in the literature with the help of data collected from human subject pedestrians and prototype implementations on commercially-available mobile and wearable devices.

\end{abstract}


\section{Introduction}
\label{intro}

\emph{Pedestrian safety} has become a critical concern in the United States (and worldwide) as the number of serious and fatal injuries due to pedestrian-related accidents continue to steadily rise every year \cite{williams2013pedestrian}. 
As one of the major causes of such pedestrian-related accidents, distracted driving has received significant attention over the past decade \cite{young2007driver,klauer2014distracted,ibrahim2011state,you2013carsafe,bergasa2014drivesafe,xu2014sober,jiang2016safecam,lin2016psafety}, which has resulted in a host of techniques to detect and overcome distraction during driving.
% , which has resulted in the enactment of various regional and national laws intended to curb distracted driving.
However, nearly 50\% \cite{bungum2005association} of all traffic related pedestrian deaths can be attributed to distraction among pedestrians (for example, inattentiveness while crossing roads and failure to obey traffic signs) rather than distracted drivers, which highlights the significant role pedestrian distraction plays in these accidents \cite{hyman2010did,schwebel2012distraction,thompson2013impact}. Besides this, distracted pedestrians are also susceptible to other non-traffic hazards in indoor and outdoor environments, such as, falling over the edge of a subway platform, walking into obstacles, falling down a stairway, colliding with other pedestrians, and falling into an uncovered sewer manhole \cite{lang2013don}. 
It is evident that distracted pedestrians pose a significant threat not only to their own safety, but also to the safety of other pedestrians (and drivers), and effective systems and mechanisms to overcome this threat are critically needed. 
%
% a  
%a generic pedestrian distraction detection and safety system, capable of accurately and efficiently recognizing a wide variety of distraction-related activities, 

%As outlined in Figure\ref{generic1},
%\begin{figure}[]
%	\includegraphics[width=\linewidth]{generic1.pdf}
%	\caption{A generic pedestrian safety framework.}
%	\label{generic1}
%\end{figure}

Designing effective pedestrian safety systems to overcome these threats, however, has been challenging. A pedestrian safety system typically comprises of two main components (Figure \ref{generic1}): (i) a \emph{distraction} or \emph{hazard detection} component, and (ii) an \emph{accident prevention} component. As any effective accident prevention strategy relies heavily on the success and timeliness of the detection component, design of efficient techniques for accurate and timely detection and recognition of pedestrian distractions and distraction-related activities is of paramount importance, and often the main technical challenge to overcome in such systems. 
%Moreover,
% which has proved to be a non-trivial challenge so far \cite{}.
%The successful usage and adoption of any pedestrian safety system or application, however, relies on the  accurate, real-time and (energy) efficient . 
The advent of sophisticated mobile and wearable devices (e.g., smartphones and smartwatches), equipped with a variety of high-precision on-board sensors capable of capturing continuous and fine-grained user movements and context, provides a tremendous opportunity to design and develop such distraction detection and recognition techniques. However, designing techniques that are fairly accurate and able to operate efficiently and in real-time, given the memory, computation and communication limitations of these devices, is not straightforward. 
%The opportunities presented by these upcoming devices, however, also come with additional challenges in the form of resource (memory and computation) constraints and communication limitations. 
%This is our primary focus in this paper. 

Several recent research efforts in the literature have attempted to improve pedestrian safety by detecting hazardous contexts (e.g., incoming vehicles, obstacles, uncovered manholes, etc.) with the help of data available from users' smartphone camera \cite{wang2012walksafe,foerster2014spareeye,wei2015automatic,peng2010smartphone} or from specialized sensors (e.g., ultrasonic sensors or depth cameras) attached to the phones \cite{hincapie2013crashalert,ahn2013casual,wen2015we}. 
%Reliance on on-board camera feed, however, limits the functionality of these scheme, for example, they do not work when the smartphone is in users' pockets and the camera feed is unavailable. Moreover, as  
Other proposals \cite{typewhilewalk,walkandtext} have employed smartphone camera to provide users with a transparent screen and a view of the road in the front, so that they can continue to safely use their phone while waking. Systems for aiding pedestrian safety that rely on other smartphone sensors such as microphone \cite{lee2011acoustic} or GPS \cite{kusano2013safety,jain2014limits,prandi2014mpass,lin2016psafety}, or on (motion) sensors on other forms of wearables such as smart footwear \cite{jain2015lookup}, have also been proposed. 
%However, these systems do not actively detect pedestrian distractions or hazards, rather they rely on the user to recognize them. 
%Also, they focus on protecting against only one particular type of distracted activity, i.e., using a smartphone while walking. 
In addition to shortcomings such as reliance on smartphone camera feed or other specialized sensors and devices which limits their functionality, several of these schemes employ computationally-intensive data processing techniques that are challenging to implement on resource-constrained mobile and wearable devices. More importantly, the above techniques fail to generalize the problem of pedestrian distraction detection by not considering a wide-variety of complex and concurrent activities that commonly resemble distraction, for example, detecting when users are walking, running or descending staircases and simultaneously reading, eating or drinking \cite{mwakalonge2015distracted,eatdistract}.
As a result, the above solutions are unable to detect or recognize a wide variety of distractions among pedestrians.

The key to designing a pedestrian safety system that has broad application and usage is to first generalize the problem of detecting distracted pedestrians as a \emph{concurrent activity recognition} (or \emph{CAR}) problem. Concurrent activity recognition is a mature technical area and results from this area have been extensively used to enable a variety of applications in health care, comfort management, personal and information security, and passive communications \cite{saguna2013complex}. Several robust and accurate CAR frameworks that detect and recognize a variety of human activities, and their complex combinations, by using data available from commercial mobile and wearable device sensors have already been proposed in the literature \cite{husz2011behavioural,korpela2015energy,krause2005trading,shoaib2016complex,liu2016complex,ma2016going,yan2012energy,shoaib2016hierarchical}. 
\emph{However, the applicability of these models for pervasive pedestrian distraction detection applications is unclear and has not been well-studied}. It appears that a majority of these CAR models proposed in the literature, owing to their use of computationally expensive data processing and analysis techniques, could be challenging to implement and/or efficiently operate on consumer-grade mobile and wearable devices with limited computational and energy resources.
Moreover, a number of other CAR systems proposed in the literature \cite{varkey2012human,borazio2013using} employ specialized/non-commercial auxiliary hardware to capture and process fine-grained user behavior, which may have cost and availability implications, thus limiting their adoption by traditional users. 

These shortcomings in existing pedestrian safety systems and concurrent activity recognition techniques necessitates investigations in two directions, and will be pursued in this paper: (i) \emph{is it possible to design a generic pedestrian distraction detection approach that can operate on existing commercial mobile and wearable devices and achieve a favorable balance between computational efficiency, detection accuracy, and energy consumption?} and (ii) \emph{how do existing concurrent activity recognition frameworks perform in a pedestrian distraction detection scenario?} Outcomes of these investigations will enable the development of a pedestrian safety system that can operate on commercially-available mobile and wearable devices and is able to accurately detect distracted pedestrian activities in a timely (real-time) fashion by utilizing reasonable device resources (processing, memory, battery). 
%The overarching goal of this paper is to improve pedestrian safety in our urban communities by contributing to the development of a pedestrian safety system which is not only effective, but also has a high chance of being adopted and used by traditional mobile users.
In line with these objectives, we first  design a novel complex activity recognition technique, called \textit{Dominant Frequency-based Activity Matching (DFAM)}, which employs a lightweight frequency matching approach on motion (accelerometer and gyroscope) data available from users' mobile and wearable devices to accurately and efficiently detect and recognize a wide variety of complex pedestrian distraction related activities. Next, we undertake a comprehensive comparative evaluation of the proposed technique with well-known complex activity recognition approaches in the literature \cite{shoaib2016complex,shoaib2016hierarchical} (specifically ones that employ classical activity classification functions, such as, \emph{k-NN}, \emph{Naive Bayes}, \emph{Decision Trees}, \emph{Random Forests}, and \emph{Support Vector Machine}) by means of data collected from real human subject pedestrians.  
In order to compare the practical feasibility and performance of these pedestrian distraction detection techniques, we also develop prototype Android and Android Wear implementations and assess their computational speed, response times and energy requirements on commercially-available Android smartphones and smartwatches under realistic operating scenarios and settings.

\section{Related Work}
\label{related}

In this section, we first outline significant mobile and/or wearable device based tools and techniques proposed in the literature for improving pedestrian safety, and discuss their limitations. 
%In the context of pedestrian safety, CAR can help in generalizing the problem of pedestrian distraction detection as the problem of recognizing a combination of pedestrian and distraction-related activities. 
As the pedestrian distraction detection problem can be generalized as a CAR problem, later we also discuss recent research results in the direction of concurrent activity recognition using these devices, primarily focusing on the recognition of human activities.
% such as gait analysis, sit-stand and stand-sit transitions, along with behavioral activities such as reading, texting, etc.

\subsection{Pedestrian Safety Systems}
%AN - What is partially sighted below?
%What is Haar?
Several research efforts in the literature have employed mobile and/or wearable devices, and data available from them, for improving pedestrian safety. \emph{WalkSafe} \cite{wang2012walksafe} utilized the rear camera of the smartphone to detect vehicles approaching a distracted user (or pedestrian) in order to promptly deliver a danger alert or notification. 
%WalkSafe is based on a machine learning algorithm utilizing Haar-like features \cite{whitehill2006haar} and is optimized with the AdaBoost algorithm \cite{freund1995desicion}.
Deng et al. \cite{wei2015automatic} used image processing techniques and multi-sensor (barometer, accelerometer and gyroscope) information on smartphones to detect surrounding objects. 
%Estimated distance and angle between the (distracted user) and surrounding objects is used to alert the user (using text, sound or vibration) about obstacles on their path. 
Similarly, Peng et al. \cite{peng2010smartphone} used real time video processing of road traffic to help partially sighted pedestrians in spotting obstacles on their path. \emph{SpareEye} \cite{foerster2014spareeye} is another proposal which applied image processing techniques on a smartphone camera feed to find obstacles in a user's path, however unlike \cite{peng2010smartphone}, SpareEye is able to track multiple obstacles simultaneously. 
One significant drawback of all these proposals is that they employ costly and resource-intensive image capture and processing techniques, which can adversely impacts the performance and battery-life of mobile devices and thus their chances of being adopted by users. 
Reliance on the smartphone's camera, also restricts the ability of these techniques to operate when the camera is obstructed, for example, in a user's pocket.

%which is able to identify only a single one obstacle at a time
% using blobs extracted from the camera feed. 
%MJ - add the below later
% \emph{UltraSee} \cite{wen2015we}  employs an ultrasonic sensor (attached to the smartphone) to detect abrupt changes in the ground surface while the user is active and using the smartphone, and thus helps user avoid potential hazards in the path.

Techniques for aiding pedestrian safety that do not rely on the camera input, but rather on a smartphone's microphone \cite{lee2011acoustic} and GPS \cite{lin2016psafety} have also been proposed. For instance, \cite{lee2011acoustic} uses sound features extracted from the smartphone's microphone to detect oncoming vehicles, while \emph{pSafety} \cite{lin2016psafety} recognizes potential collisions between pedestrians and oncoming vehicles using the smartphone's GPS. One major drawback of these systems is that they are useful in detecting only outdoor traffic-related hazards scenarios.
%in the pedestrian path, and lacks ability to differentiate between a passing vehicle or an approaching vehicle.  and notifies them both through mobile networks. But this framework can be , not indoors. 
Furthermore, techniques that employ specialized devices and sensors for improving pedestrian safety have also been proposed. \emph{Lookup} \cite{jain2015lookup} uses information from specialized motion sensors attached to pedestrians' shoes to profile step and slope in order to detect curbs, ramps and other obstructions. Similarly, Ramos and Irani \cite{hincapie2013crashalert} used a depth camera (paired with a smartphone), while Ahn and Kim \cite{ahn2013casual} and \cite{wen2015we} employed an ultrasonic sensor for detecting pedestrian hazards and/or for guided navigation.
% for identifying obstacles in a pedestrian's path. Similarly,  used  for  through pedestrian traffic.
Besides relying on specialized sensors, these systems attempt to address pedestrian safety by detecting obstacles or other potential hazards (to pedestrians). In this paper, we take an orthogonal approach to pedestrian safety by attempting to detect distraction or inattentiveness among pedestrians; after all if pedestrians are not distracted they will be able to easily navigate away from obstacles and other hazards (including, traffic). 
%do not attempt to e a pedestrian's current activity or context which can put them in harm's way.

% these frameworks are impractical as they require users to purchase additional equipment and also increase the overall weight of the hardware setup. 
%But these solutions focus on improving pedestrian safety by detecting obstacles or potential hazards in a user's (or pedestrian's) path rather than the user distraction itself, and hence we require new pedestrian safety solutions capable of recognizing both potential hazards, and distractions. %While such applications can be helpful in certain scenarios, continuous detection and notifications (even when users are alert) could easily overwhelm users \cite{garzonis2008mobile,westermann2015assessing}. This is especially true for urban users, where there are several objects and/or other pedestrians in the user's surroundings.  For example,  It is possible that these mobile networks may be overwhelmed with alerts especially in an urban(crowded) setting, where there pedestrians and drivers count in the vicinity are high. If users are overwhelmed with alert notifications, they may ignore critical alerts, or even stop using the application \cite{aranda2016m}.As a result, it is more desirable to have \textit{a system that recognizes when (pedestrian) users are distracted, and notifies them of potential dangers if and only if they are recognized to be distracted}.

\subsection{Concurrent Activity Recognition (CAR)}
The problem of detecting distracted pedestrians can be generalized as a concurrent activity recognition or CAR problem where the goal is to detect concurrent pedestrian activities of being mobile (e.g., walking, running or climbing/descending stairs) and being distracted (e.g., texting, eating or reading).
% activities can be recognized by employing models that can be designed to . 
%The most significant advantage of 
CAR techniques that can distinguish different combinations of elementary activities have been extensively used in the literature for complex human activity recognition. For instance, Shoaib et al. \cite{shoaib2016complex} used \textit{multi-source} and \textit{multi-sensor motion} data, from two smartphones, one in trouser pocket and the other on the wrist, to recognize activities that involve hand gestures, such as smoking, eating, drinking coffee and giving a talk. 
%Their evaluation results show that the combination of these sensor placement positions, wrist and thigh, outperforms either of them used in isolation.  
Liu et al. \cite{liu2016complex} also employed multi-sensor time series data to recognize sequential, concurrent, and generic complex activities by building a dictionary of time series patterns (called \emph{shapelets}) to represent atomic activities.
%perform CAR. Their framework . They then present three shapelet-based models.
In another related effort, Shoaib et al. \cite{shoaib2016hierarchical} proposed a multi-layer approach to detect non periodic concurrent activities such as smoking while walking, where the first layer employs a traditional classifier (e.g., Random forest and SVM) for activity identification while the second layer incorporates context rules to correct mis-classifications. Alternatively, Husz et al. \cite{husz2011behavioural} and Ma et al. \cite{ma2016going} proposed techniques to recognize concurrent and complex actions from camera feed or video data.
%Husz et al. \cite{husz2011behavioural} modeled independent tracking and behavioral analysis of activities, to allow recognition of concurrent and complex actions, from motion captured in video data. Similarly, Ma et al. \cite{ma2016going} explored the use of deep convolution neural networks to recognize actions, objects and activities, using camera feed.
%MJ-what is the drawback of \cite{shoaib2016complex}?
However, several shortcomings in these approaches prevent them from being effectively used in pedestrian safety applications. For instance, \cite{shoaib2016hierarchical} requires the system to keep track of time segments that precede and follow the current one and thus unsuitable for pedestrian safety applications that require real-time operation and feedback. Others are not suitable for implementation on resource-constrained mobile and wearable devices, primarily due to their use of complex feature sets and classification functions (as in \cite{liu2016complex,ma2016going}) or resource-intensive image capture and processing techniques (as in \cite{husz2011behavioural}).
%, and is therefore unsuited for solutions that usually require instantaneous notifications to the pedestrian for avoiding plausible dangers. 
%Additionally, most of the techniques and systems \cite{husz2011behavioural,liu2016complex,ma2016going} . Specifically, the use of complex feature sets and classification functions (as in \cite{liu2016complex,ma2016going}) and continuous image capturing and processing techniques (as in \cite{husz2011behavioural}) can be very resource intensive, and adversely impact the performance and battery life of users' mobile/wearable devices. This, in turn, diminishes the feasibility of its implementation on wearable devices and its overall adoption by end users. 
As discussed before, one of the main functional requirement for a mobile/wearable device based CAR framework for pedestrian safety is computational and energy efficiency. Earlier research efforts in energy-aware recognition mechanisms \cite{krause2005trading,yan2012energy} have achieved a favorable balance between classification accuracy and energy consumption, but these schemes have been successful in recognizing only simple activities, such as, standing, walking and sitting, but not concurrent (and distracted) activities. Recently, Korpela et al. \cite{korpela2015energy} proposed an energy-aware CAR framework for real-time applications by using a minimal feature set to recognize individual data segments and an hierarchical classification mechanism for concurrent activity recognition.  However, their framework employs a specialized wearable device hardware, and may not work for commercial off-the-shelf mobile devices.

\section{Pedestrian Distraction Detection}
\label{system}

\begin{figure}[b]
%\centering
	\includegraphics[width=\linewidth]{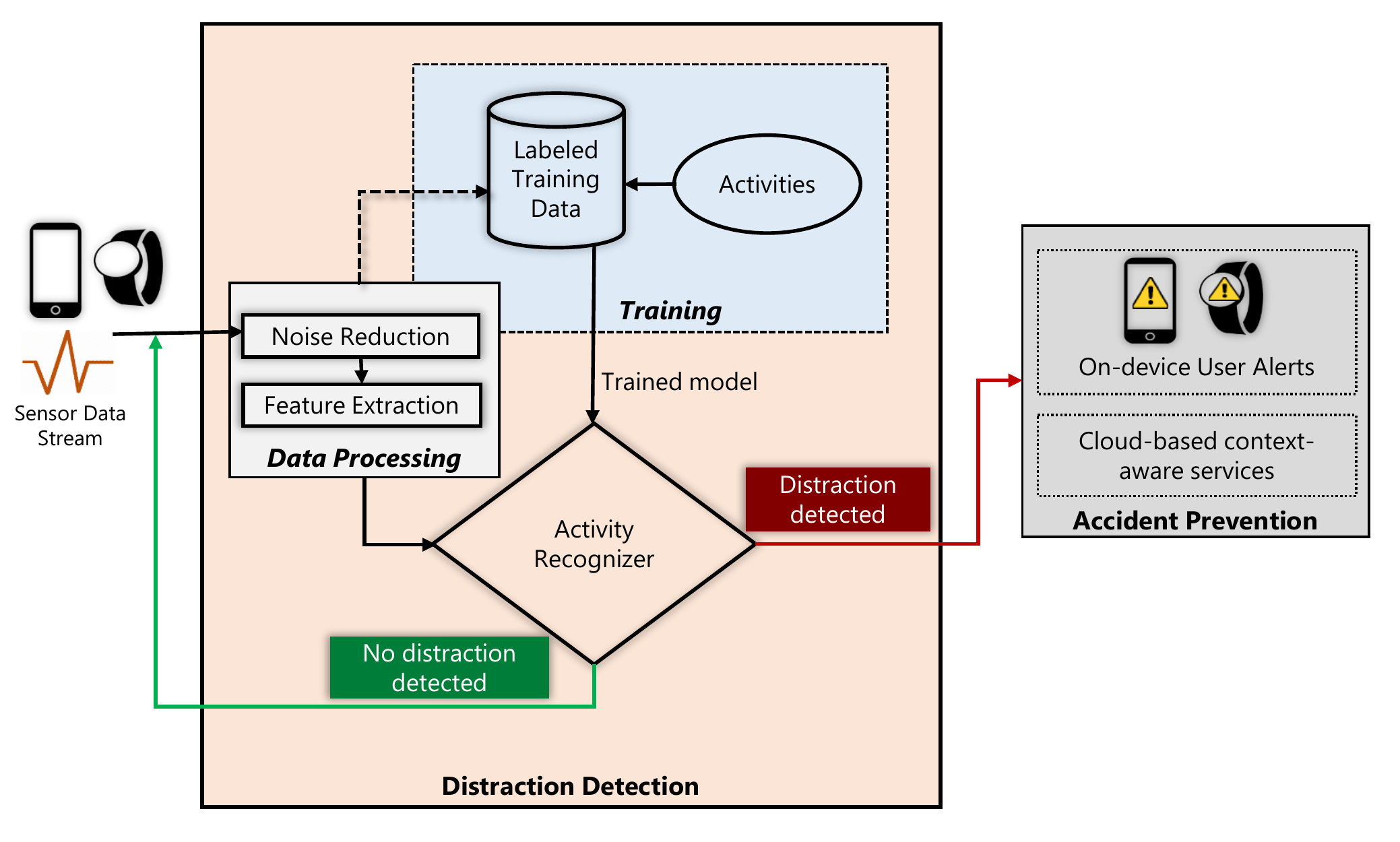}
	\caption{A generic pedestrian safety system.}
	\label{generic1}
\end{figure}

As outlined earlier, any pedestrian safety system typically comprises of two main components (Figure \ref{generic1}): (i) a \emph{distraction} or \emph{hazard detection} component, and (ii) an \emph{accident prevention} component. In this paper, we primarily focus on the former.
% due to its criticality to the system as a whole.
Figure \ref{generic1} depicts the design of a generalized learning based framework which is the main building block for pedestrian distraction detection in such systems. As shown in the figure, the distraction detection framework comprises of: (i) a \emph{data processing module} (includes, noise removal, segmentation, and feature generation), and (ii) a \emph{CAR model building phase} (includes, design of an appropriate activity classification function and training it using processed and labeled training data). Once a trained CAR model is available, it can be employed to recognize (or classify) distracted pedestrian activities. 
%recognition using the trained CAR models.
Such a design of the distraction detection framework is commonly employed in the literature (and in practice) for pedestrian safety and other applications, and will also be employed by us in this paper. 
Our distraction detection framework relies on multi-sensor data obtainable from multiple mobile devices carried by the pedestrians, specifically, motion (including, data from both accelerometer and gyroscope sensors) and contextual information from the pedestrian's smartphone and smartwatch.
% (ii) motion sensor (includes both accelerometer and gyroscope) data from the smartwatch donned by the pedestrian, and (iii) context sensitive information (system flags) from the smartphone. 
The data processing module in our framework filters this multi-sensor data (to eliminate errors and inconsistencies), segments it into fixed-size blocks or windows, and extracts relevant features from it. Features extracted from a set of labeled  training data are then used to train appropriate CAR models within a supervised learning paradigm. It should, however, be noted that the data processing task (including, features and feature extraction) may vary depending on the chosen CAR technique. These trained CAR models are then utilized by our framework for distracted activity classification or recognition tasks. As an appropriate CAR technique is central to the design of a pedestrian distraction detection framework that can attain a practicable balance between computational efficiency, detection accuracy and energy consumption, in this section we focus on designing such a technique. Technical details of our proposed CAR technique, referred to as DFAM, are presented next. We also outline details of other well-known techniques that have been employed in the literature \cite{shoaib2016complex,shoaib2016hierarchical} for similar activity classification tasks, as we later empirically compare the performance of our DFAM technique against these classical activity classification techniques.

%\subsection{Distracted Activity Recognition}
%To recognize  activities, we design an efficient CAR model that employs frequency domain analysis of the motion sensor data, and compare it (in Section \ref{evaluation}) with five other well-known parametric classification techniques generally used for CAR, namely, Naive Bayes, k-NN, Decision Trees, Support Vector Machine and Random Forests \cite{shoaib2016complex,shoaib2016hierarchical} .In this section, we first discuss the technical details of the CAR classification function DFbAM, and  later examine how five other well-known classification functions in the literature can also be used in our framework in place of DFbAM.

\subsubsection{DFAM} 
Our DFAM CAR technique is inspired from the audio matching algorithm proposed by Avery Wang \cite{wang2003industrial}. Proprietary versions of Wang's algorithm is commonly used in popular song searching applications, such as \emph{Shazam}. However, it is not trivial to directly use Wang's audio matching algorithm for activity recognition using motion data. First, there are significant differences between audio data (found in audio files) versus motion data (sampled from the smartphone and smartwatches). In the audio matching application (or song searching, as in Shazam), matching features in the test audio file occur at almost identical relative time offsets from the beginning of the audio file being matched to. 
In contrast, motion data from pedestrian activities generally do not occur at exactly fixed time offsets, therefore requiring a new matching algorithm. Other differences between motion and audio data include a significantly lower sampling rate of smartphone and smartwatch motion sensors (compared to audio data which is generally sampled at a much higher frequency) and distinctly different \emph{dominant frequency} ranges of both types of data. 
Recently, Sharma et al. \cite{sharma2008frequency} successfully applied \emph{dominant frequency-based activity matching} for simple (non-concurrent) activities, using fixed threshold-based classifiers. In this paper, we use pre-processing techniques used by Wang \cite{wang2003industrial} and extend Sharma et al.'s work significantly, in order to recognize concurrent activities related to pedestrian distractions in our framework.

\textbf{DFAM Training:}
\label{dft}
%DFAM utilizes dominant frequency analysis on 
During the training phase, (low-pass) filtered time-series motion data from the smartphone and smartwatch, denoted as $T_p$ and $T_w$ respectively, corresponding to each activity of interest is first \emph{segmented} into smaller fixed-sized windows of $W$ samples. Let's assume that this motion data is sampled at a frequency $f_s$. 
% each of $T_p$ and $T_w$ 
 %In other words, a labeled training dataset $T_p$ and $T_w$ is represented as:
\begin{equation}
T_p = \{\textsuperscript{1}b_p, \textsuperscript{2}b_p, \ldots, \textsuperscript{m}b_p\} \textrm{;} \quad T_w = \{\textsuperscript{1}b_w, \textsuperscript{2}b_w, \ldots, \textsuperscript{n}b_w\}
\end{equation}
\begin{equation*}
\textrm{where} \quad m = \frac{sizeof(T_p)}{W} \quad \textrm{and} \quad n = \frac{sizeof(T_w)}{W}
\end{equation*}
After this pre-processing step, the frequency response of each window in $T_p$ and $T_w$ is independently calculated using a discrete Fourier transformation technique such as a fast Fourier transform (or FFT \cite{batenkov2005fast}). Let the frequency responses corresponding to $T_p$ and $T_w$ be represented as $F_p$ and $F_w$, respectively.
\begin{equation}
F_p = \{\textsuperscript{1}r_p, \textsuperscript{2}r_p, \ldots, \textsuperscript{m}r_p\} \textrm{;} \quad F_w = \{\textsuperscript{1}r_w, \textsuperscript{2}r_w, \ldots, \textsuperscript{n}r_w\}
\end{equation}
\begin{equation*}
\textrm{where} \quad \textsuperscript{i}r_p = FFT(\textsuperscript{i}b_p) \quad \textrm{and} \quad \textsuperscript{i}r_w = FFT(\textsuperscript{i}b_w)
\end{equation*}
Each of the frequency response blocks $\textsuperscript{i}r_p \in F_p$ and $\textsuperscript{i}r_w \in F_w$ are then analyzed for a dominant frequency in $g$ (empirically determined) frequency bins\footnote{The highest possible frequency component is $\frac{f_s}{2}$ \cite{zayed1993advances}.}, with one dominant frequency in each bin:
\begin{equation*}
\{f_{(0,u_1)}, f_{(u_1,u_2)}, \dots, f_{(u_{g-1},\frac{f_s}{2})}\} 
\end{equation*}
All of the observed dominant frequency in each of the $g$ bins are then compressed or hashed to create a `\emph{signature}' for the activity. 
%MJ-Check with Nisha and Anindya and write a sentence about the hash/compression function
As we are employing multiple devices and sensors, with each sensor possibly outputting measurements across multiple dimensions (e.g., each accelerometer sensor measurement is across three dimensions), each training data point will consist of measurements across multiple dimensions. For example, a dominant frequency analysis on three-dimensional ($x, y, z$) time series data window will result in a three-dimensional training point $\langle H_x, H_y, H_z\rangle$, where $H_x$, $H_y$, and $H_z$ are the hashes of dominant frequencies on respective axes. Now, let us denote the set of all distracted activities as $\mathbb{D}$, and the set of all pedestrian activities as $\mathbb{P}$. For each activity $a_u \in\mathbb{P}$, a training dataset made of equalized data points is created during the training phase, and stored along with the corresponding label $a_u$. Similarly, for the each concurrent activity $a_v \in\mathbb{P}\times\mathbb{D}$, another training dataset made of equalized data points is created during the training phase, and stored along with the corresponding label $a_{v}$.
%AN - You use $a_i$ for both distracted and pedestrian activities. This can be confusing.

\textbf{DFAM Activity Classification:}
To correctly classify the current or test user activity (say, $a_{c}$), DFAM employs a dominant frequency matching technique using the labeled training data (from the previous phase), as described below.
%) by employing  obtained in the training phase. 
Given a test window with $s$-axis signatures, the activity is matched using the following scoring function:
\[ S_{i,j}(a_{c}) =
  \begin{cases}
    0       & \quad \text{if } \sum_{k=1}^{s} F(\textsuperscript{c}H_k , \textsuperscript{train}H^{i,j}_k) = 0\\
    (\frac{1}{s})^s  & \quad \text{if } \sum_{k=1}^{s} F(\textsuperscript{c}H_k , \textsuperscript{train}H^{i,j}_k) = 1\\
    (\frac{2}{s})^s  & \quad \text{if } \sum_{k=1}^{s} F(\textsuperscript{c}H_k , \textsuperscript{train}H^{i,j}_k) = 2\\
    \quad \vdots  & \quad \quad \quad \quad \quad \quad \quad \vdots\\
    (\frac{s-1}{s})^s  & \quad \text{if } \sum_{k=1}^{s} F(\textsuperscript{c}H_k , \textsuperscript{train}H^{i,j}_k) = s-1\\
    1  & \quad \text{if } \sum_{k=1}^{s} F(\textsuperscript{c}H_k , \textsuperscript{train}H^{i,j}_k = s\\
  \end{cases}
\]

where $S_{i,j}(a_{c})$ is the matching score per training instance $j$ in each activity $a_i\in\mathbb{P}\times\mathbb{D}$, $\textsuperscript{c}H_k$ is the current activity signature from $k$-th sensor axis, $\textsuperscript{train}H^{i,j}_k$ is the signature from $k$-th sensor axis of $j$-th training instance of activity $a_i$, and 
\[ F(a,b) =
  \begin{cases}
    0 & a \neq b\\
    1 & a = b\\
  \end{cases}
\]
The above scoring function gives exponentially more weight to multi-dimensional signature matches, which will intuitively result in a higher score when matching with the ground truth activity. Finally, the activity is classified after matching against the entire training dataset of all activities as follows:
\begin{equation}
\\ arg\max_{i}  \sum\limits_{j} S_{i,j}(a_{c}) \quad \forall a_i \in \mathbb{P}\times\mathbb{D}
\label{egclass}
\end{equation}
The current activity $a_{c}$ is then classified as that activity $a_i$ which achieves the maximum aggregated score as shown in Equation \ref{egclass}.
%In order to identify activities outside the cartesian product \mathbb{P}\times\mathbb{D}, we propose imposing a threshold on the final scores. If the final classified activity score falls below this threshold, then the test activity does not fall under the domain \mathbb{P}\times\mathbb{D}.

\subsubsection{Traditional Classifiers}
\label{traditionalclassifiers}
Traditional supervised learning-based classification functions, such as, Naive Bayes, k-NN, Decision Trees, Support Vector Machine and Random Forests have been successfully used in the literature (and in practice) for detecting complex and concurrent human activities \cite{shoaib2016complex, shoaib2016hierarchical}. Given that distracted pedestrian activities are inherently concurrent activities, these supervised learning based techniques comprise of a suitable candidate set for a comparative performance evaluation with our proposed DFAM technique. Below, we outline how these classification techniques are employed within our pedestrian distraction detection framework, and provide details on the related data pre-processing, feature extraction and model training tasks.   

%MJ- Can the number of features and window sizes be dynamic parameters??
\textbf{Data Processing:} 
The (low-pass) filtered time-series motion data from the smartphone and smartwatch, denoted as $T_p$ and $T_w$ respectively, corresponding to each activity of interest is first \emph{segmented} into smaller fixed-sized windows, as discussed earlier for DFAM. Like before, each of the motion data stream $T_p$ and $T_w$ comprises of both the accelerometer and gyroscope sensor data sampled along all the three axes at some frequency $f_s$.
%We first \emph{clean} the raw sensor data streams obtained from both the smartphone ($T_p$) and smartwatch ($T_w$) before processing the data. During the cleaning process, we remove any redundant and noisy data samples from $T_p$ and $T_w$ using a low-pass filter. We then \emph{segment} the cleaned data into smaller blocks ,
A set of time and frequency domain \emph{features} are then computed from each window of the time-series motion data streams, which are widely used in the literature for activity recognition \cite{shoaib2016complex,sun2010activity,varkey2012human,parate2014risq,ilmjarv2015detecting}.
%, denoted as X = {$x_1$, $x_2$, $…$, $x_l$},
%block we \emph{extract}   where $x_i$ represents a . 
The computed features for each window include:
% from the accelerometer and the gyroscope sensor three-dimensional data ($x$, $y$, $z$) and they  
%Need to reduce number of features and mention why they were chosen
\begin{itemize}
\item Mean, minimum, maximum, standard deviation, variance, along with energy and entropy of discrete FFT components for each of the three axes of both the accelerometer and gyroscope time-series data.  
\item Root mean square (RMS) correlation measures among the three axes for each of the accelerometer and gyroscope time-series data.
\item Mean, median, and maximum of the instantaneous speed (only for the accelerometer data)
\item Mean, median, and maximum of roll velocity (only for the gyroscope data)
\end{itemize}  
As the motion time-series of an activity comprises of several windows, features computed for all the windows are combined to create a feature set for that activity.
This process (filtering, segmentation, and feature extraction) is repeated for all the considered distraction-related activities in $\mathbb{D}$ and non-distraction activities in $\mathbb{P}$ in the training dataset to create a labeled feature set for all the activities. Such a labeled training (feature) set is then used to train each of the concurrent activity classification models, details of which are outlined below. 
%we use WEKA \cite{hall2009weka} to train the CAR model with the block feature set X, and activity label (in case of supervised or label-based learning), and employ the trained model to classify or recognize an unknown activity(block data) with feature set Y.
%Need to mention that we use existing tool to do the training..WEKA.
It should be noted that the above data pre-processing and feature extraction tasks remain the same for all the supervised learning based classification functions considered below.
%  procedures are synonymous; they differ only in the way they are trained. 

\textbf{Naive Bayes (NB):} Given a test (or unlabeled) feature set, a trained NB model estimates the posterior probabilities of each activity (in the set of all considered concurrent activities $\mathbb{P}\times\mathbb{D}$), 
%classes for the data sample, based on the conditional independence 
assuming that the input features are independent of each other. The unknown or test activity is then assigned an activity label corresponding to the maximum posterior probability value. Posterior probabilities are estimated according to the Bayes rule as follows:
\begin{equation}
P(a|X)= \frac{P(a) P(X|a)}{P(X)}
\end{equation}
where, $P(a|X)$ is the posterior probability of the activity $a$ given a feature set $X$, $P(a)$ is the prior of the activity $a$, $P(X|a)$ is the likelihood of the feature set given $a$, and $P(X)$ is the probability of occurrence of the feature set which is independent of the activity.

\textbf{Decision or Classification Tree (DT):} This technique constructs a tree structure using the labeled feature sets of the training data, where leaves of the tree represent the class labels of the different concurrent activities, whereas the branches represent conjunction of features that lead to these class labels.
%models generate binary trees based on hierarchical, conditional branching using labelled training data samples, where each branch leads to nodes which either split into branches(intermediate nodes) or do not(leaf nodes). Each intermediate node represents a condition on a feature $x_i$, whereas the leaf nodes represent an activity class or label.
Now given the feature set of an unknown activity, the corresponding activity label is determined by traversing through the branches of the trained tree model using the discrete feature values in the feature set until a leaf node is reached. The unknown activity is then classified with the label corresponding to the reached leaf node.

\textbf{Random Forests (RF):} These are ensembles of decision trees that could output multiple activity labels, one each from a decision tree, for an unknown activity sample. 
The unknown activity is then assigned an activity label using a majority rule.

\textbf{Support Vector Machine (SVM):} This technique uses the labeled training data (feature sets) to learn the hyperplanes separating the different activity classes. These hyperplanes or decision boundaries are optimized to achieve maximum separation distance between activity classes. After the model is trained, i.e. separating hyperplanes are determined, the feature set corresponding to an unknown activity is assigned a label corresponding to its placement in the bounded n-dimensional feature space.

\textbf{k-Nearest Neighbours (k-NN):} This technique creates a trained classification model by grouping labeled training data or feature sets into separate clusters based on their class label or the activity they represent.
% clusters or groups together input data samples(blocks) from the same activity class, where each cluster in the trained model represents a different class(activity). 
Then, data sample from an unknown activity (or an unlabeled feature set) is classified as the class (or activity) of a majority of its {k} closest or nearest neighbors. 
%by first choosing {k} most closest neighbors and utilizing the majority voting approach after determining the test sample's proximity to them.
%The representative cluster for the  Y) is identified from the trained model(clusters), by  
Euclidean distance can be used as a closeness measure to compute the proximity between two feature sets.

\section{ Evaluation and Results}
\label{evaluation}

In this section, we evaluate the performance of the DFAM and other activity recognition schemes as CAR techniques used for detecting distracted pedestrian activities. 

\subsection{Experimental Setup}
We collect motion sensor data of distracted pedestrian activities using a wrist-worn smartwatch and a paired smartphone (Motorola Moto XT1096). To test the versatility of our framework, we test it across two different smartwatches, namely a Sony Smartwatch 3 and a LG Urbane W150. A combination of smartwatch and smartphone was placed on participating pedestrians\footnote{A total of 23  participants took part in our study, which was approved by Wichita State University's Institutional Review Board (IRB).}, for a total of four different device placement scenarios. For the same-side placements, either both smartwatch and smartphone are worn on the right wrist and placed inside right hip pocket (RR), or worn on the left wrist and placed inside the left hip pocket (LL). The remaining two scenarios alternate the placements to the opposite sides, i.e., smartwatch on right wrist along with phone in left hip pocket (RL), smartwatch on left with phone in right pocket (LR). Each participant performed a pre-defined but randomized set of activities for one or more of the scenarios. The set of activities consisted of non-pedestrian, pedestrian and distracted-pedestrian related activities outlined in Table \ref{activitylist}. All concurrent activities except the starred (*) activities form a set of distracted pedestrian activities.

\begin{table}[h]
\fontsize{6}{6}\selectfont
\centering
\caption{Activities performed by the participants.}
\label{activitylist}
    \begin{tabular}{|p{1.6cm}|p{3.25cm}p{2.45cm}|}
    \toprule
    Simple Activities & Concurrent Activities&\\
    \midrule
    Standing  & Walking + Using Smartphone & Walking + Reading\\
    Walking & Climbing stairs + Eating & Walking + Eating\\
    Climbing stairs & Descending stairs + Eating & Walking + Drinking\\
    Descending stairs & Climbing stairs + Drinking & Standing + Drinking*\\
    Sitting & Climbing stairs + Using Smartphone  & Standing + Reading*\\
    Running & Descending stairs + Using Smartphone & Standing + Eating*\\
     & Running + Using Smartphone  & Sitting + Using Smartphone*\\
     & Standing + Using Smartphone* & Descending stairs + Reading\\
     & Descending stairs + Drinking & Climbing stairs + Reading\\
    \bottomrule
    \end{tabular}
\end{table}

We developed a custom Android application using Android Studio IDE v2.2.3 running on Java 8 platform, to record activity related motion sensor data from the Moto T1096 running on Android 6.0 and the smartwatches running on Android Wear 1.5, at a sampling rate of 50 Hz. The activity data collected includes three-dimensional accelerometer and gyroscope sensor data from the aforementioned devices. Throughout the data collection, we took several precautionary measures to ensure participant safety during certain distracted activities, due to potential falling and injury risks. For example, we placed a safety harness on the participant when descending stairs and reading at the same time. However, we also ensured that these safety measures did not interfere with the activities performed by the participants. For more details on how individual activities were carried out, please refer to \cite{expdetails}. On an average, each participant took about 2 hours to complete all the activities. The physical demands of our experiments, together with these additional constraints in selecting participants thereby limited our ability to recruit a larger number of participants, or obtain data for all possible device placements from the same participant. 

We implemented the proposed framework using Java on (i) a 64-bit Debian Linux PC with an Intel Core i5 processor and 8 GB RAM and (ii) the Motorola Moto XT1096 smarphone. Implementations for the traditional classifiers were derived from the Weka 3 machine learning toolkit \cite{hall2009weka} and its Android counterparts. The PC implementation was used to extensively analyze the performance of DFAM, which is not possible on a resource constrained smartphone. On the other hand, the smartphone implementation is helpful in evaluating on-device response time and resource utilization in real-life usage. Next, we explore how different training sets can affect the classification accuracy of the proposed DFAM scheme. This step is crucial for the pedestrian safety application, because not all users would be willing to setup a personally trained model. In other words, personalized datasets may not be a realistic scenario, and thus the proposed DFAM scheme should work well with unseen test sets.

\subsection{DFAM Performance}
\label{dfamperformance}

\begin{figure*}[]
\begin{subfigure}{0.49\linewidth}
\centering
\includegraphics[width=\textwidth]{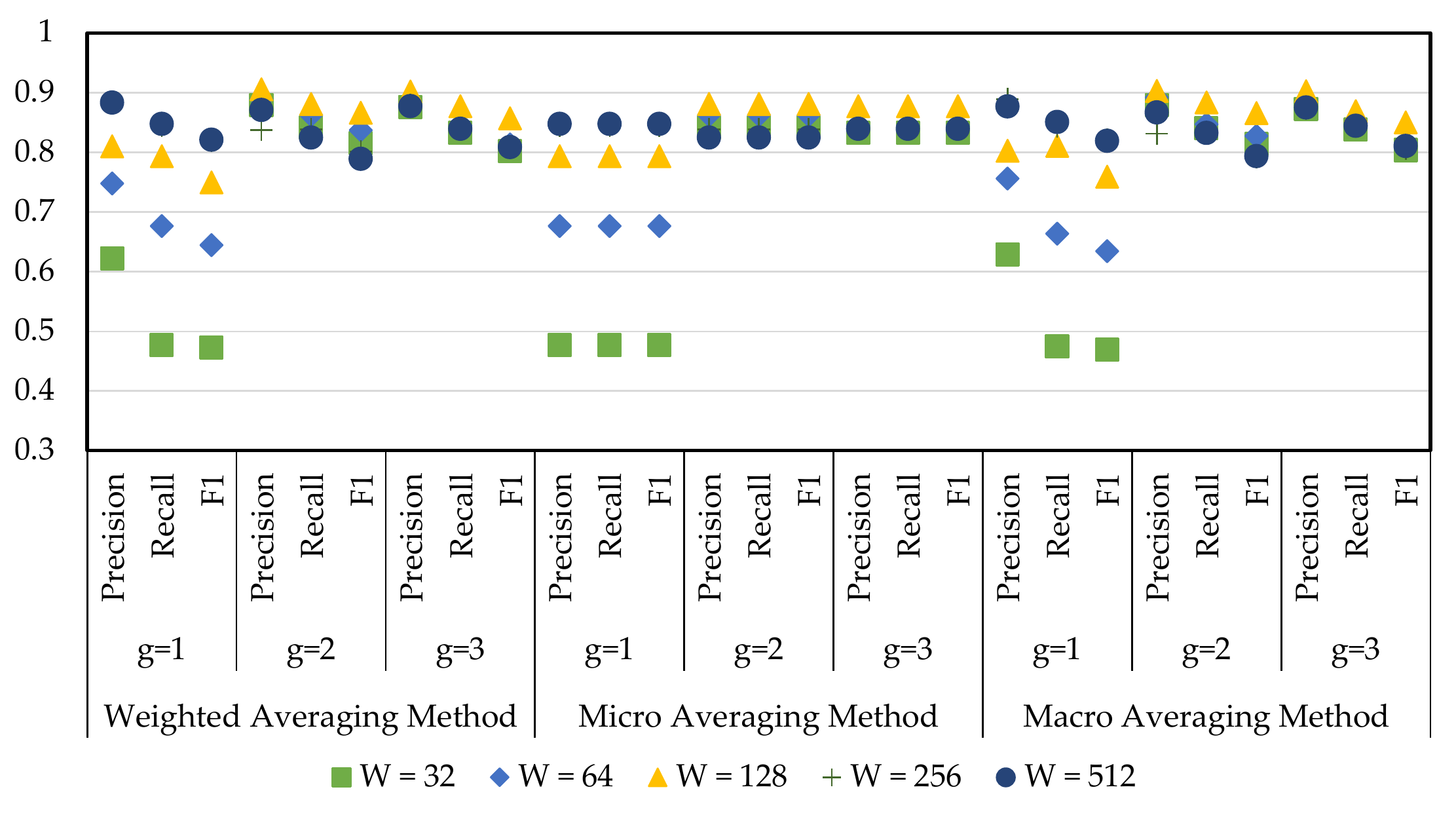}
\caption{}
\label{dfam-sony-same}
\end{subfigure}
\hfill
\begin{subfigure}{0.49\linewidth}
\centering
\includegraphics[width=\textwidth]{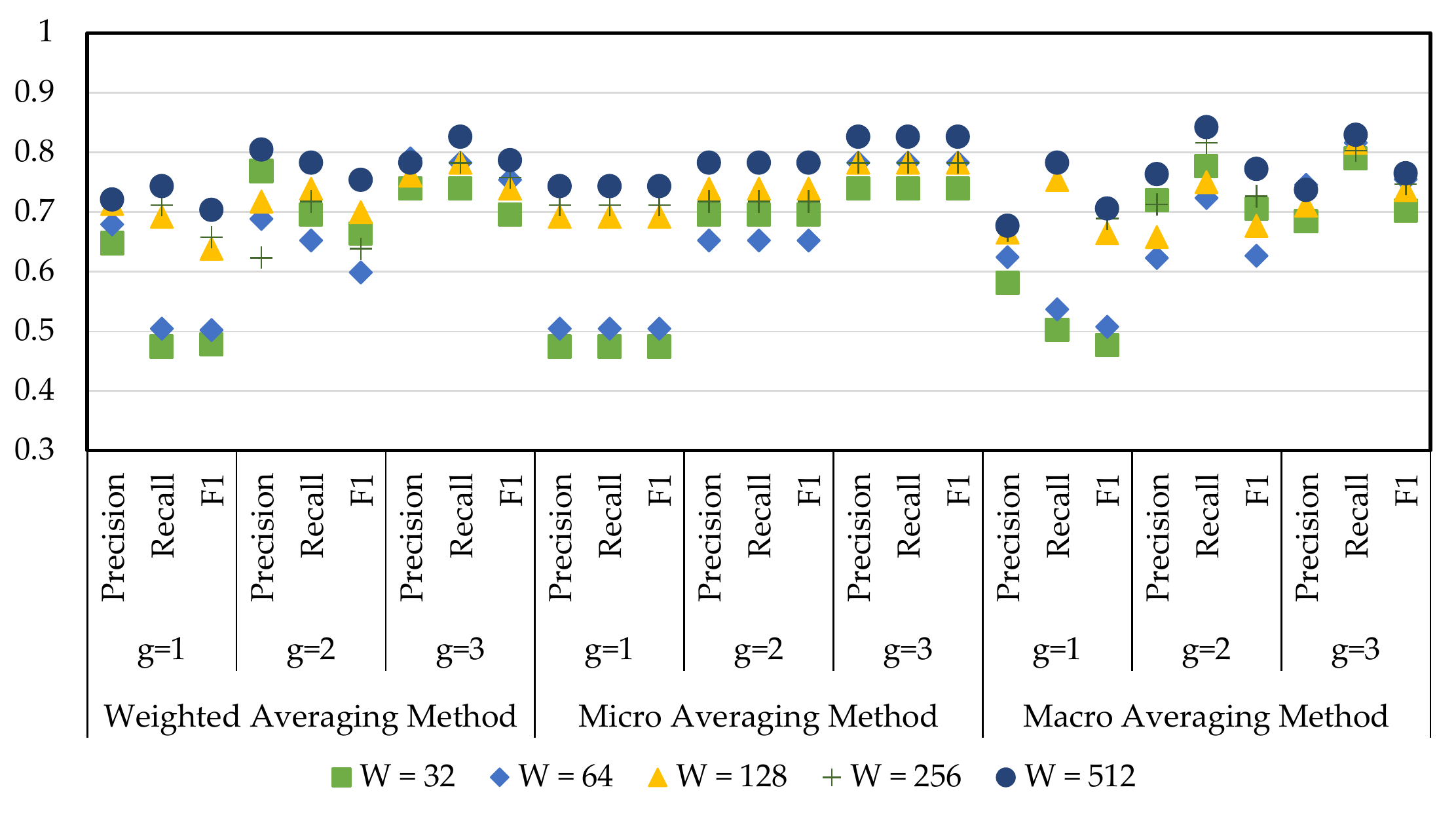}
\caption{}
\label{dfam-lg-same}
\end{subfigure}
\hfill
\begin{subfigure}{0.49\linewidth}
\centering
\includegraphics[width=\textwidth]{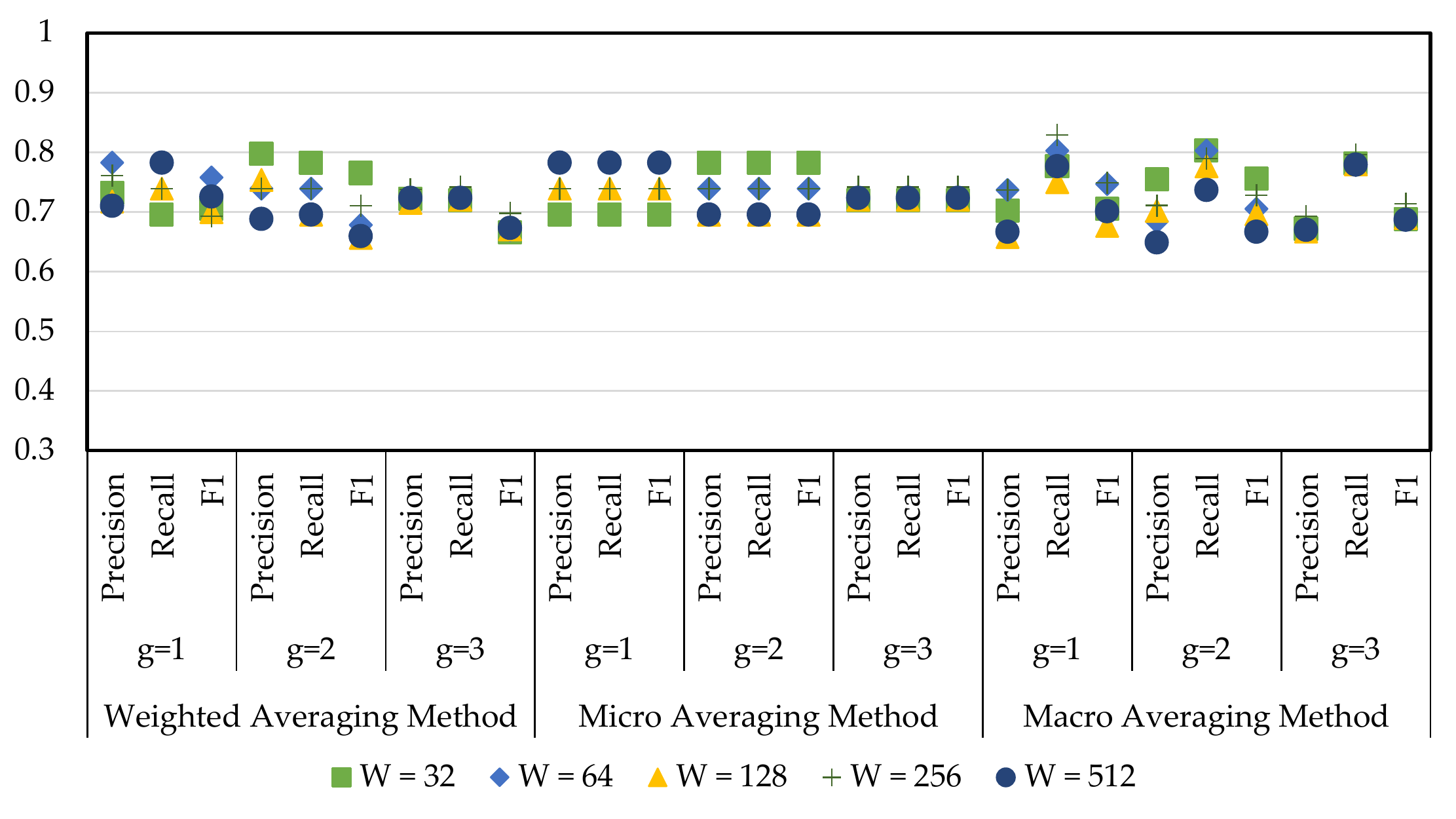}
\caption{}
\label{dfam-sony-diff}
\end{subfigure}
\hfill
\begin{subfigure}{0.49\linewidth}
\centering
\includegraphics[width=\textwidth]{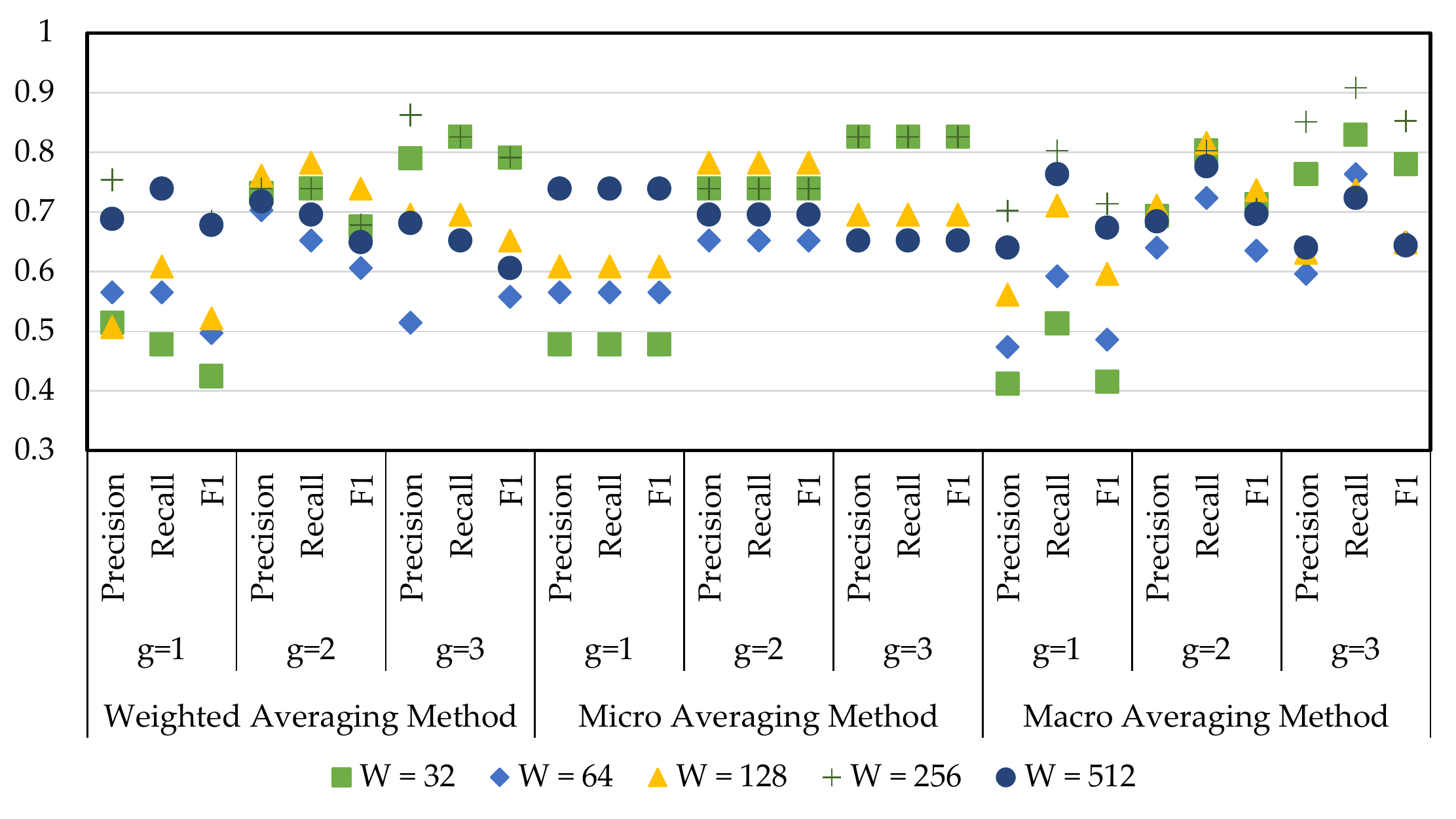}
\caption{}
\label{dfam-lg-diff}
\end{subfigure}
\caption{DFAM performace for datasets of (a) same-side using Sony+Moto, (b) same-side using LG+Moto, (c) different-side using Sony+Moto, and (d) different-side using LG+Moto.}
\label{dfam-devices-placement}
\end{figure*}

\begin{figure}[t]
\begin{subfigure}{0.495\linewidth}
\centering
\includegraphics[width=\textwidth]{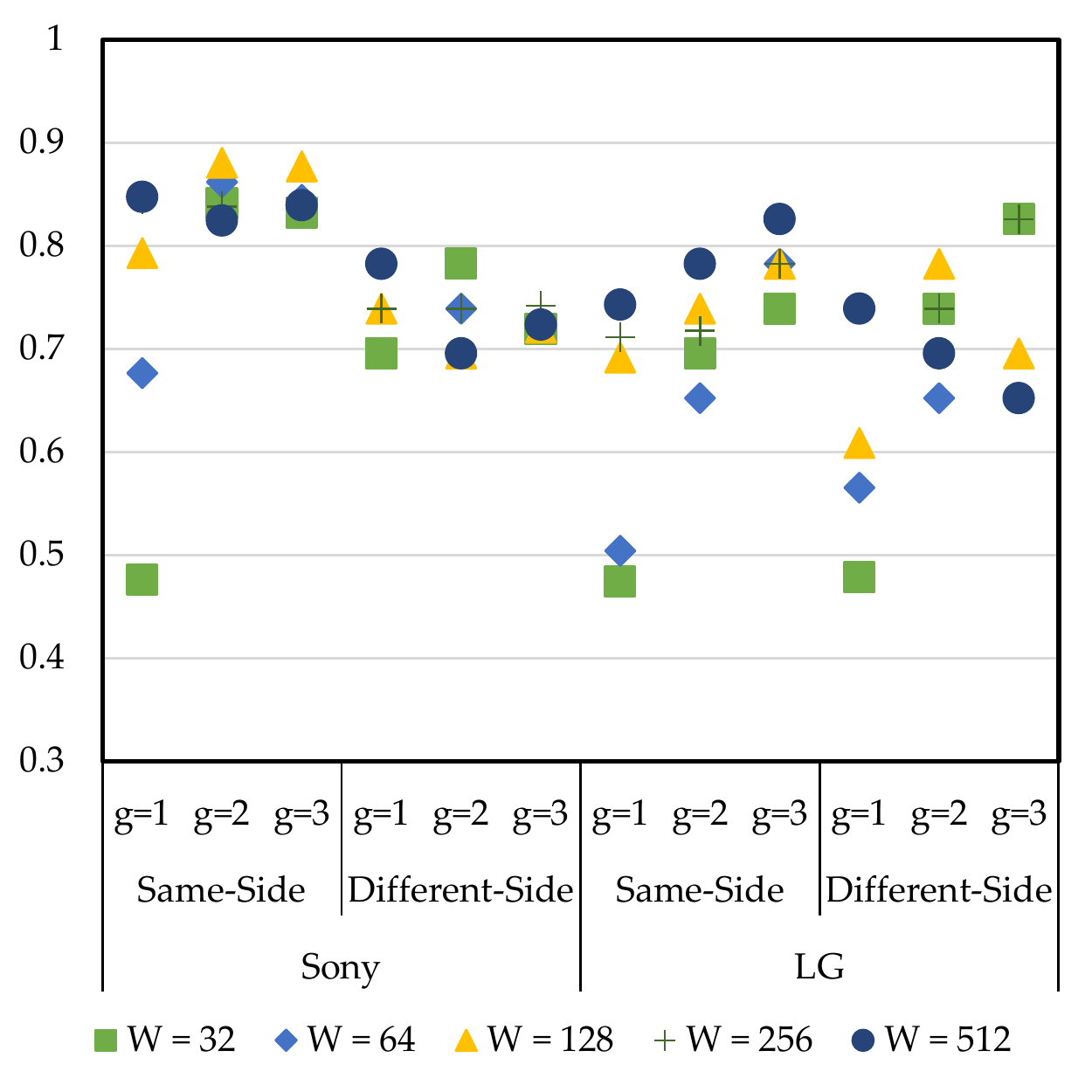}
\caption{}
\label{dfam-ca}
\end{subfigure}
\hfill
\begin{subfigure}{0.495\linewidth}
\centering
\includegraphics[width=\textwidth]{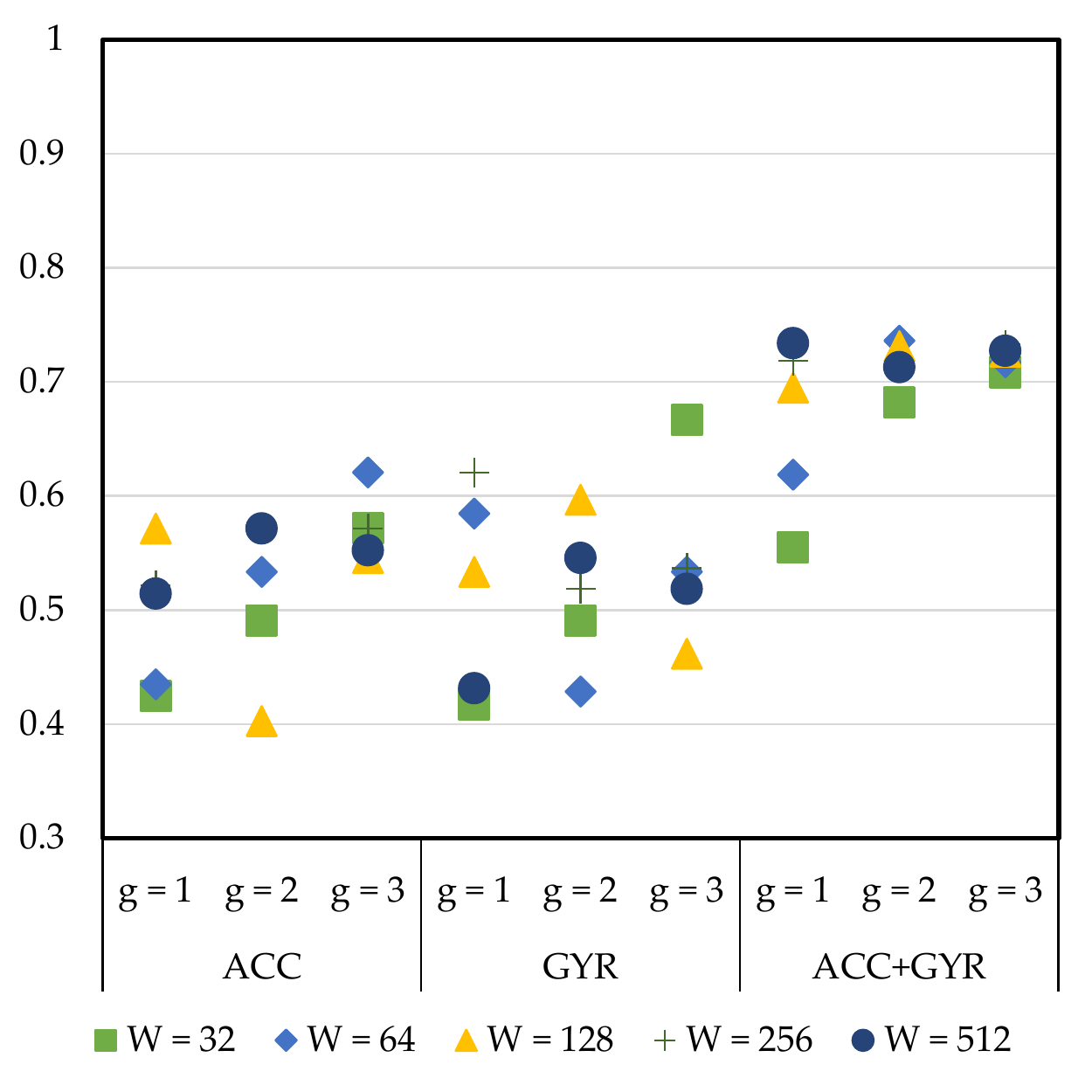}
\caption{}
\label{dfam-sensors}
\end{subfigure}
\caption{Classification accuracy of DFAM for (a) combined datasets, and (b) individual sensors.}
\end{figure}

We first validate the feasibility of detecting distracted pedestrian using DFAM, by creating personalized models for each participant using their individual datasets, and then performing a Leave-One-Out Cross Validation (LOOCV) using the trained model. In LOOCV, one block is allotted as test data, while the rest remain in the training set. The individual participant accuracies are then averaged out for each group, where the datasets are grouped based on the smartwatch used, and further grouped based on the device placements (Table \ref{dfam-datasets}). For different window $W$ and bin sizes ($g$) of the collected data, we evaluate DFAM performance across three different averaging methods -- weighted, micro and macro -- based on metrics such as classification accuracy, precision, recall and F1 score as shown in Figures \ref{dfam-devices-placement} and \ref{dfam-ca}.

\begin{table}[h]
  \centering
  \scriptsize
  \caption{Datasets collected per placement scenario.}
    \begin{tabular}{|r|rrrr|r|}
    \toprule
     & RR & LL & RL & LR & Total \\
    \midrule
    LG+Moto & 5     & 6     & 4     & 4     & 19 \\
    Sony+Moto & 6     & 6     & 5     & 5     & 22 \\
    \midrule
    Total & 11    & 12    & 9     & 9     & 41 \\
    \bottomrule
    \end{tabular}%
  \label{dfam-datasets}%
\end{table}%

Figures \ref{dfam-sony-same} and \ref{dfam-lg-same} show the precision, recall and F1 scores for same-side (LL, RR) device placement, where as figures \ref{dfam-sony-diff} and \ref{dfam-lg-diff} show the precision, recall and F1 scores for different-side (LR, RL) device placement. We observed that the precision and recall improve with increasing number of frequency bins. The mean F1 score for $g=3$ was 0.7475, compared to 0.6586 and 0.7335 for $g=1, 2$, respectively, for all four placement scenarios combined. We did not observe any significant performance difference using weighted, micro and macro averaging methods between the Sony+Moto and LG+Moto datasets. The mean classification accuracy (for $g=3$ and $W=\{32, 64, 128, 256, 512\}$) for Sony+Moto and LG+Moto datasets are 0.79 and 0.75, respectively, with a standard deviation of 0.07 and 0.06, respectively. This implies that the proposed DFAM is implementable across different wrist-based wearables. We also observed slight performance difference between the same-side and different-side datasets. The mean classification accuracy (for $g=3$ and $W=\{32, 64, 128, 256, 512\}$) for same-side and different-side datasets are 0.81 and 0.72, respectively, with a standard deviation of 0.04 and 0.05, respectively. This implies that DFAM works slightly better for same-side smartphone and smartwatch placement.

Next, we investigate the effect of combining datasets from different participants from the same group to obtain a trained model, and validate it using $k$-fold cross validation where $k$ equals 10. In 10-fold cross validation, the dataset is split into 10 equal parts, one of which becomes the testing set, and the remaining nine folds constitute the training set. We compute the classification accuracies for different window $W$ and bin $g$ sizes as shown in Figure \ref{dfam-ca}. DFAM achieved classification accuracy of 0.70 for $g=3$ and $W=32$ (0.64 seconds at 50 Hz). As intuitively expected, classification accuracy improves as the window size is increased to $W=512$ (10.24 seconds at 50 Hz), although at a cost of increased detection time as we evaluate later in Section \ref{real-resource}. 

In a real world implementation, it may not be practical to combine data exclusively from participants having the same hardware. Moreover, not all wearables may have both accelerometer and gyroscope sensors, compelling us to examine whether DFAM can classify the activities in the absence of either gyroscope (GYR) or accelerometer (ACC) data as shown in Figure \ref{dfam-sensors}.  We observed that classification accuracy dropped to 0.57 when using only accelerometer data, for $g=3$ and $W=32$ (compared to 0.70 in ACC+GYR datasets). Similarly, classification accuracy was 0.66 when using only gyroscope data, for $g=3$ and $W=32$. This implies that DFAM works better in the presence of both accelerometer and gyroscope sensors. We also reinforce our earlier observation that classification accuracy is highest for $g=3$. As a result, we set bin size $g=3$ for all following experiments.

\subsection{Comparison with Traditional Classifiers}

A realistic setting involves using already trained models to recognize activities of a previously unseen participant. We evaluate and compare DFAM in such a setting by leaving out one participant's dataset for testing purposes and training using the rest. This Leave-One-Subject-Out (LOSO) approach validates the generalization performance of the CAR schemes. We compare the classification accuracies of DFAM, and other CAR schemes for different window sizes ($W=\{32, 64, 128, 256, 512\}$) as shown in Table \ref{dfamvsothers}. Results show that DFAM's classification accuracy is comparable with SVM, DT, RF, NB and 1-NN. However, 2-NN and 3-NN performs better than DFAM in most cases, but they also impose higher resource utilization as we evaluate next in Section \ref{real-resource}.

\begin{table}[]
\centering
\fontsize{5.49}{6}\selectfont
\caption{Classification accuracy of DFAM compared with traditional classifiers.}
    \begin{tabular}{|r|rrrrrrrr|}
    \toprule
          & DFAM & SVM   & DT    & RF    & NB    & 1-NN   & 2-NN   & 3-NN \\
    \midrule
    W = 32 & 0.4538 & 0.4298 & 0.4744 & 0.5063 & 0.5243 & 0.5203 & 0.4733 & 0.5133 \\
    W = 64 & 0.4845 & 0.4408 & 0.572 & 0.58  & 0.464 & 0.637 & 0.52  & 0.644 \\
    W = 128 & 0.5396 & 0.4774 & 0.4182 & 0.507 & 0.482 & 0.5234 & 0.5242 & 0.5218 \\
    W = 256 & 0.5099 & 0.4803 & 0.4892 & 0.6718 & 0.5379 & 0.5114 & 0.5224 & 0.515 \\
    W = 512 & 0.5392 & 0.5424 & 0.5367 & 0.5227 & 0.5421 & 0.5373 & 0.5409 & 0.532 \\
    \bottomrule
    \end{tabular}
  \label{dfamvsothers}
\end{table}

\begin{table*}[]
\centering
\caption{Average response time and resource utilization of DFAM compared with traditional classifiers.}
    \begin{tabular}{|r|rrrrrrrr|}
    \toprule
              & DFAM & SVM   & DT    & RF    & NB    & 1-NN   & 2-NN   & 3-NN \\
    \midrule
    Response Time (ms) & 640-1150 & 2000-6000 & 1200-4480 & 2800-10630 & 890-2980 & 2800-8600 & 1900-8600 & 2900-6600 \\
    CPU Utilization & 0.5-4.5\% & 1.3-8\% & 0.3-1.6\% & 0.5-7.4\% & 0.7-2.4\% & 1.5-3.8\% & 1.2-3\% & 1.2-2.9\% \\
    Power Consumption (mW) & 33.3-129.5  & 33.3-188.7 & 33.3-85.1 & 85.1-222 & 40.7-96.2 & 85.1-214.6 & 85.1-188.7 & 85.1-218.3 \\
    RAM Utilization (MB) & 20-24 & 26-53 & 17-67 & 56-108 & 19-29 & 15-26 & 29-53 & 30-92 MB \\
    \bottomrule
    \end{tabular}
\label{real-resource-table}
\end{table*}

\subsection{Response Time and Resource Utilization}
\label{real-resource}
We next evaluate the response time, CPU, RAM and power consumption of the CAR models on the Motorola XT1096 smartphone paired with the Sony Smartwatch 3. The XT1096 with a 2300mAh Li-ion battery was running Android 6.0, while the Smartwatch 3 with a 420mAh Li-ion battery was running Android Wear 1.5. For this analysis, we use the same participant data and pre-trained classification models from Section \ref{dfamperformance}. However, in this case the signature (or feature) generation and matching (or classification) is executed on the mobile and wearable device, unlike the previous evaluation (Section \ref{dfamperformance}) where they were executed on a PC. Table \ref{real-resource-table} compares the response time and resource utilization of DFAM compared with traditional classifiers, with ranges signifying varying window sizes  ($W=\{32, 64, 128, 256, 512\}$). The response time excludes the communication delays, time taken to obtain a data block and consumed time not related to generation of block-related features, which would be same for all the techniques. The time taken to obtain the data blocks remained constant across the different CAR models for a window size $W$, along with the time taken to generate features across the traditional CAR models. The CPU utilization, power consumption and RAM utilization were also recorded in these trials over a period of two minutes and repeated 10 times. The RAM usage is measured in megabytes (MB), whereas the CPU utilization is in percentage indicating the fraction of available processing power used.

Results show that DFAM has significantly lower response time compared to traditional classifiers, which is beneficial for alerting distracted pedestrian in real-time. CPU utilization, power consumption and RAM utilization are also on the lower side for DFAM, which means users will notice minimal impact on performance of their smartphone. Notably, 2-NN and 3-NN which achieved slightly better classification accuracy earlier, also have the highest response times and generally consumes more system resources. Quick response time is vital in determining the effectiveness of our framework, because any delay in alerting distracted pedestrians can be decisive in potential accident preventions. These results positions DFAM as the preferred candidate for use as a CAR technique in the distracted pedestrian detection framework.

\subsection{A Hierarchical CAR for Distracted Pedestrian Detection} 

\begin{figure}[t]
\centering
\includegraphics[width=0.86\linewidth]{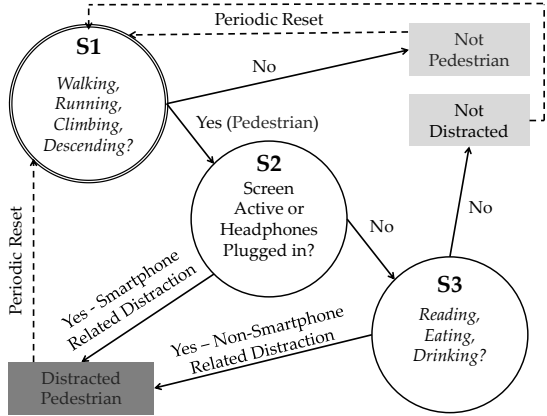}
\caption{Hierarchical CAR using DFAM.}
\label{dfam-har}
\end{figure}

We propose a modification to the CAR model presented earlier in order to improve detection accuracy, with minimal impact on resource utilization and response time. We propose to use a \emph{hierarchical CAR} model representing various user activities in form of states (Figure \ref{dfam-har}), where simple or concurrent activity classification is performed only within the corresponding states. In state S1, only simple pedestrian activities are observed using smartphone and smartwatch motion data. Once a pedestrian activity is detected, the flow moves to state S2, where contextual information is used to determine two possible types of distractions: \emph{smartphone-related distractions} and \emph{non-smartphone-related distractions}. For smartphone-related distractions the detection and notification is relative straightforward, whereas non-smartphone-related distractions are observed using both smartphone and smartwatch motion data (S3). If a pedestrian is observed to be distracted, appropriate notification is given, and the flow moves back to S1 after a periodic reset. Such a hierarchical approach will reduce the computational and communication overheads involved in activity recognition, because the CAR model is executed only within a subset of all the states (S1 and S3 in Figure \ref{dfam-har}). As a result, the smartphone can preserve system resources during non-pedestrian and smartphone-related distraction activities. The hierarchical approach can also improve the individual classification accuracies of states S1 and S3, because the classification problem in each of these states is reduced from a multi-class classification problem to a binary classification problem. 

For implementing the improved hierarchical framework, we modified the on-device Android application by splitting the activity set into simple and concurrent activity sets for S1 and S3, respectively, and trained a DFAM model for each. The simple pedestrian activity is determined first (S1 activities), followed by the distracted or concurrent activity which is triggered only when the pedestrian is moving (S3 activities), and not when the pedestrian is using the smartphone (S2 conditions). Another modification to the application involves building a service to synchronize and trigger sensor data logging on the smartwatch only when the smartphone starts logging sensor data. This allows the smartwatch to go into sleep mode, and consequently save energy (by approximately 33.3 mW) when the concurrent activity recognition state is not triggered. The hierarchical approach achieved between 91-96\% classification accuracy for pedestrian activities (S1) and 77-89\% for non-smartphone related-distraction activities (S3), with a slightly decreased CPU utilization (0.4-1.4\%). This range of percentages for the classification accuracy and CPU utilization are for an increasing value of the window sizes $W=\{32, .., 512\}$.

\section{Discussions and Future Work}
\label{discussions}

\textbf{Accident Prevention:} In this work, we limit ourselves to solely study the viability of using a CAR based framework for improving pedestrian safety. In order to prevent unwanted distraction related injuries and fatalities, we plan to develop an \textit{on-device alert module} for users' mobile and/or wearable devices, to remind distracted pedestrians that they should pay more attention to their surroundings while they are in motion. Additionally, we plan to implement a \textit{cloud-based alert module} that will employ crowd-sourced contextual information from distracted users to alert other users in the vicinity about the presence of distracted pedestrians. 
The design of these alert modules, however, is not trivial and requires a careful analysis of the associated human-factors issues. An alert mechanism that is not carefully designed may annoy users with frequent notifications, who may in turn decide not use it anymore, or may become a source of distraction themselves. We plan to accomplish this as part of our future work.

\textbf{Number of Bins:} In this work, we observed that the classification accuracy of DFAM improves as the number of bins $g$ increases. It is possible that this trend continues for higher bin sizes, i.e., $g>3$. However, having $g>3$ for smaller window sizes did not yield good classification accuracies (as seen in Figure \ref{dfam-devices-placement}), thus making it difficult to do an equitable and meaningful analysis along side larger window sizes. Also, the resource utilization may change for more number of bins. As part of our future work, we will focus on empirically determining the optimal number of bins ($g$) for each window size that can achieve an improved trade-off between classification accuracy and resource utilization.

\textbf{Other Future Work:} 
%Multiple other challenges also need to be addressed before such an assistive technology can be effective and widely adopted. 
%For example, only if a significant percentage of the population is using our framework, and voluntarily reporting their distraction conditions, the cloud-based context-aware alerts will become effective. 
%Privacy and security concerns may also hinder the adoption of such context-aware crowd-sourced applications. In future, we plan to develop privacy-preserving schemes that can facilitate such assistive applications without requiring the user to reveal their activities. The battery consumption needs to be further reduced in order to have an imperceptible footprint.
In this work, we validated the performance of the proposed CAR technique (DFAM) and its Android implementation across different wearable device hardwares, i.e., smartwatches. It will also be interesting to study how the performance of the proposed CAR technique and its implementation varies across different smartphone hardwares. Also, we reckon that other CAR techniques (e.g., k-NN, RF, DT, NB or SVM) can also be employed in the hierarchical CAR model. However, the viability and efficiency of these techniques would depend on the feature set they use (for classification) in each state of the hierarchical model. As part of future work, we plan to conduct a comprehensive comparative evaluation of the hierarchical model comprising of these other techniques.
 
\section{Conclusion}
\label{conclusion}

We outlined and comprehensively evaluated a novel framework that detects and recognizes distracted pedestrian activities by using motion data available from users' mobile and wearable devices. As part of our framework, we designed and evaluated a novel dominant frequency matching based concurrent activity recognition model, called DFAM, and compared the performance and execution efficiency of the DFAM model with other well-known learning-based classification functions, such as Random Forests, SVM,  k-NN, Naive Bayes and Decision Trees. Our evaluation results showed that the proposed DFAM model is a suitable candidate for detecting concurrent activities, such as that of distracted pedestrians, and that it has reasonable concurrent activity recognition accuracy compared to traditional classification functions, such as, k-NN, Naive Bayes and Decision Trees. We also observed that DFAM has lower power consumption rates and quicker response time(s) compared to Random Forests, SVM,  k-NN, Naive Bayes and Decision Trees. In summary, we have not only comprehensively evaluated the efficacy and feasibility of various concurrent activity recognition techniques for detecting and recognizing pedestrian distraction, but have also proposed a novel concurrent activity recognition technique that achieves a good balance between recognition accuracy and alert response time, while being energy efficient.

\section*{Acknowledgments}
This research work was supported by the National Science Foundation (NSF) under award number CNS-1637290.

\bibliographystyle{IEEETran}
\bibliography{sigproc}
\end{document}